\documentclass[sn-mathphys,Numbered]{sn-jnl}

\usepackage{graphicx}%
\usepackage{multirow}%
\usepackage{amsmath,amssymb,amsfonts}%
\usepackage{amsthm}%
\usepackage{mathrsfs}%
\usepackage[title]{appendix}%
\usepackage{xcolor}%
\usepackage{textcomp}%
\usepackage{manyfoot}%
\usepackage{booktabs}%
\usepackage{algorithm}%
\usepackage{algorithmicx}%
\usepackage{algpseudocode}%
\usepackage{listings}%
\usepackage{soul}
\usepackage{fix-cm}

\begin{document}

\title[Article Title]{Estimating the link budget of satellite-based Quantum Key Distribution (QKD) for uplink transmission through the atmosphere}


\author[1]{\fnm{Satya Ranjan} \sur{Behera}}

\author*[1]{\fnm{Urbasi} \sur{Sinha}}\email{usinha@rri.res.in}


\affil[1]{{Raman Research Institute}, \orgaddress{\street{C.V. Raman Avenue, Sadashivanagar}, \city{Bengaluru}, \postcode{560080}, \state{Karnataka}, \country{India}}}

\abstract{Satellite-based quantum communications including quantum key distribution (QKD) represent one of the most promising approaches toward global-scale quantum communications. To determine the viability of transmitting quantum signals through the atmosphere, it is essential to conduct atmospheric simulations for both uplink and downlink quantum communications. In the case of the uplink scenario, the initial phase of the beam's propagation involves interaction with the atmosphere, making simulation particularly critical. To analyze the atmosphere over the Indian subcontinent, we begin by validating our approach by utilizing atmospheric data obtained from the experiments carried out in the Canary Islands within the framework of Quantum Communication (QC). We also verify our simulation methodology by reproducing simulation outcomes from diverse Canadian locations, taking into account both uplink and downlink scenarios in Low Earth Orbit (LEO). In this manuscript, we explore the practicality of utilizing three different ground station locations in India for uplink-based QC, while also considering beacon signals for both uplink and downlink scenarios. The atmospheric conditions of various geographical regions in India are simulated, and a dedicated link budget analysis is performed for each location, specifically focusing on three renowned observatories: IAO Hanle, Aries Nainital, and Mount Abu. The analysis involves computing the overall losses of the signal and beacon beams. The findings indicate that the IAO Hanle site is a more suitable choice for uplink-based QC when compared to the other two sites.}




\maketitle

\section{Introduction}\label{sec1}

Towards the late 1980s, Bennett, Brassard, and their colleagues developed an experimental prototype in the IBM lab, enabling the successful implementation of a quantum key exchange. This was the first instance of free space QC being used for key exchange with the transmission distance limited to 320 mm \,\cite{bennett_experimental_1992,smolin_early_2004,brassard_brief_2006}. Between 1990 and 2000, more free space QC has been conducted, with a 1-kilometer maximum distance \,\cite{jacobs_quantum_1996,hughes_quantum_2000,buttler_daylight_2000,resch_distributing_2005}.
In 2002, the National University of Singapore and the University of Vienna conducted free-space experiments, resulting in the successful distribution of entanglement over a distance of 7.8 kilometers during night-time \,\cite{erven_entangled_2008}. 
When the distance between two locations is limited to a few kilometers \,\cite{13km,7km,23km}, the impact of the atmosphere on QC is minimal. However, as the distance extends to tens of kilometers, the atmosphere becomes increasingly influential in QC. Yet, these studies do not offer detailed analyses of the contribution of the atmosphere to the signal. A noteworthy milestone was reached in 2007 with the collaboration of the European Space Agency (ESA) and SECOQC, where a significant quantum communication link spanning 144 km between the Canary Islands of La Palma and Tenerife was successfully established \,\cite{ursin_entanglement-based_2007, PhysRevLett.98.010504,144km}. Due to the 144 km line of sight distance, the atmosphere played a significant role in this experiment. The communication channel involved in this study is horizontal. This experiment was conducted in actual atmospheric conditions, featuring channel attenuation similar to that encountered in an optical link between the ground and a low Earth orbit satellite. Therefore, it certainly demonstrates the viability of a technologically comparatively straightforward method for satellite-based quantum key distribution.

Fiber-based communication experiences considerable loss when the transmission distance is beyond the line of sight. To overcome this limitation, satellite-based techniques can be executed through uplink or downlink setups. In 2016, China successfully launched the world's first quantum communication satellite, marking a significant milestone in global quantum communication efforts 
 \,\cite{chen_integrated_2021,liao_satellite--ground_2017,dai_towards_2020,yin_entanglement-based_2020}. This was a downlink communication system. In a downlink setup, communication involves a transmitter situated on the satellite and a ground-based receiver. This arrangement offers advantages like a high key generation rate due to reduced photon loss and minimal dark counts. Furthermore, it remains notably resistant to atmospheric turbulence. However, it's worth noting that placing the transmitter on the satellite contributes to an increase in the overall payload weight, potentially leading to escalated project complexity and costs.

On the contrary, the uplink configuration lacks constraints regarding photon source quantity or weight. It adds more flexibility to any change of the photon source required in the future, which is possible with the source on the ground. Also, it will allow for the inclusion of supplementary optical components, such as those required for error correction and characterization. However, in the case of uplink, the beam encounters the atmosphere in the initial stage of its propagation. The effect of the atmosphere on attenuating the link and reducing the key rate is unavoidable. Hence, atmospheric simulation plays an important role in modelling uplink compared to downlink-based QC.

In the realm of free-space quantum communication, a thorough grasp of atmospheric influences is required. These effects encompass phenomena like beam wandering, beam diffusion, scintillation, pointing error, and link attenuation as the transmitted beam navigates through a turbulent medium. This comprehension stands as a prerequisite for achieving effective communication between the satellite and the ground station. Various strengths of turbulence and their effect on quantum signals have been explained in \,\cite{ghalaii_quantum_2022}. A comprehensive numerical simulation was conducted, utilizing realistic simulated orbits and accounting for factors such as pointing error, diffraction, atmospheric conditions, and telescope design \,\cite{Bourgoin_2014} for several geographic locations in Canada. This simulation aimed to provide estimates of the losses that a satellite-based system would encounter.

The Canary Island experiments \cite{ursin_entanglement-based_2007, PhysRevLett.98.010504,144km} present experimental results, indicating complete losses where turbulence accounts for 18 dB, atmospheric transmittance contributes 15 dB, and the remaining losses are attributed to optical factors, resulting in a total loss of 40 dB. The initial step involves simulating these outcomes with all the parameters to validate the loss values. The procedure's specifics are outlined in sections A and A.1 of the appendix. This horizontal free-space communication spanning 144 kilometers exhibits losses that are equivalent to those experienced in the LEO. In fact, simulations indicate that the anticipated link transmittance from an LEO satellite will be quite similar \cite{aspelmeyer_long-distance_2003}.
In contrast, the Canadian study \,\cite{Bourgoin_2014}, contains simulation findings of satellite loss in LEO orbit at various locations across Canada. To demonstrate that our simulation is accurate, we extended it to the LEO orbit and compared the findings with those from Canada. The detailed verification is given in A.2 of the Appendix.
After comparing our simulation strategy with experimental results \cite{ursin_entanglement-based_2007, PhysRevLett.98.010504,144km} and simulations conducted in other geographical regions \,\cite{Bourgoin_2014}, we proceed to apply it in detail to diverse scenarios within India. Specifically, our focus is on three distinct regions for three well-known observatories in the near-IR and optical bands. One of the observatories is the Indian Astronomical Observatory (IAO) in Hanle which is the highest astronomical observatory in India at 4500 meters altitude in the Himalayan region \,\cite{surendran_development_2018}. Situated in the Shivalik mountain range, the Aryabhatta Research Institute of Observational Sciences (ARIES) in Nainital serves as the second observatory \,\cite{chen_turbulence_2022}. ARIES is 1951 meters above sea level, and during the observation, the monsoon plays a crucial role. The third observatory is near Mount- Abu, which is located close to the Thar Desert at 1680 meters above sea level, where there is almost no rain for most of the year. The varied atmospheric conditions make India a fascinating study for varied geographical regions and their applicability to space-based QKD.

The manuscript is organized as follows: In Section 2, the loss budget between the satellite receiver and the ground station is calculated, whereas the same calculation is done for the beacon signal in Section 3. The key rate of two separate sources, i.e., weak coherent pulses (WCP) and entangled sources, is computed in Section 4 of the manuscript. Section 5 is the Conclusion section, and the final Section is the Appendix carrying additional information.
 
\section{Link Budget analysis}
One of the first steps in building a line-of-sight communication system is creating a link budget, which serves many important purposes, including performance estimation before the link is formed, determining, often for a specified worst-case scenario, whether there is enough optical power to cross the link.
Table\ref{tab:LinkBudget} shows an example of the link budget calculation. For this, we have chosen IAO Hanle as the ground station location, and the satellite orbit at 500 km above the ground station at low earth orbit (LEO). The signal wavelength we have considered is 810 nm. The details of the parameters used for the link budget are discussed later in this section.  

If $P_{t}$ is the transmitted power then the detector aboard the satellite receives power $P_{r}$, which is given by \cite{toyoshima_optimum_2002}.
 
\begin{equation}
     P_{r}=P_{0}IS\,,
\end{equation}
   
$P_{0}$ is the received power in the absence of a turbulent atmosphere. It is assumed that the highest obtainable power is $P_{0}$, at the satellite. The next two parameters $I$ and $S$ are due to the atmosphere. $I$  has a beta distribution due to atmosphere and pointing-related jitter, while $S$ normally has a lognormal distribution as a result of scintillation \,\cite{Kiasaleh1994OnTP}.

In the absence of air turbulence, the received power at the satellite's detector is primarily affected by free-space propagation loss and optical absorption. When accounting for optical losses at both the transmitter and receiver, as well as atmospheric absorption, the resultant received power can be calculated as follows \,\cite{NEL} 

\begin{equation}
    P_{0} = P_{t}\eta^{\sec\theta}_{atm} L_{r}\eta_{t}\eta_{r}G_{t}G_{r}\,,
\end{equation}
where $G_{t}$, which is the gain of the transmitter, $G_{r}$ is the receiver telescope gain and $L_{r}$ is the Free space propagation loss given by,
\begin{equation}\label{eqgt}
    G_{t} = \frac{8}{\Theta^2_{B}}\,, \ G_{r} = \frac{4\pi A_{r}}{\lambda^2} \ \text{and}\ L_{r} = (\frac{\lambda}{4\pi L})^2\,.
\end{equation}
  ${\Theta_{B}}$ is the divergence of the laser beam in radians and $L$ is the communication range in meters. 

   \begin{itemize}
   \item $\eta_{t}$ is the optical efficiency of the transmitter.
   \item $\eta_{r}$ is the optical efficiency of the receiver.
    \item $ A_{r}$ represents the aperture area of the receiver, expressed in square meters.
    \item $\eta^{\sec\theta}_{atm}$ is the atmospheric attenuation at zenith.
    \end{itemize}
Table \ref{tab:LinkBudget} gives the link budget calculation. The signal beam is chosen with a 20$\mu rad$ divergence. This can be achieved with additional beam-expanding optics before the telescope. This small divergence keeps the beam tightly collimated over the 500km path, resulting in a compact footprint at the satellite. A smaller spot size significantly increases photon-collection efficiency at the receiver. However, the beam with this lower divergence also raises pointing error loss, meaning even tiny misalignments can cause a large fraction of the beam to miss the receiver’s aperture.

\begin{center}
\begin{table}[!ht]
\centering
\begin{tabular}{|c|c|c|c|}
\hline 
No & Parameter & unit & signal   \\
\hline 
1 & Tx gain  ($ G_{t}$) & dB & 109.03  \\
2 & Tx Beam divergence ($2 \Theta_{B}$) & $\mu rad$ & 20\\
3 & Tx optics loss ($ \eta_{t}$) & dB & -2.20 \\
\hline
4 & Path loss ($L_{r}$) &  dB & -257.79 \\
5 & Atmospheric attenuation ($\eta_{atm}$)=0.651 & dB & -1.84  \\
6 & Beam Wander loss $(L_{BW})$ & dB & -0.40  \\
\hline 
7 & RX gain ($G_{r}$) & dB &  121.32  \\
8 & RX optics loss ($ \eta_{r}$) & dB & -2.2 \\
9 & RX pointing loss & dB &  -1.83 \\
\hline 
10 & Total loss & dB & 35.91  \\
\hline 

\end{tabular}
\caption{Link budget calculation for the IAO Hanle observatory. Row 1, 2, and 3 represent parameters for the transmitter telescope. Row 4, 5 and 6 represent propagation losses. Row 7, 8, and 9 represent parameters for the receiver telescope.}
\label{tab:LinkBudget}
\end{table}
\par\end{center}

The link budget for an optical system can be categorized into three main components: (i) Transmitter gains, which typically involve telescopes, and (ii) losses in optical path propagation. It's in two parts, losses through vacuum space and through the atmosphere, and finally (iii) the losses experienced by the received signal as it travels through the optical receiving components to reach the detector \,\cite{chan_optical_2000}.

\subsection{Transmitter:}
 A telescope converts the source's quantum beam into a signal that is sent to the receiving satellite.  The beam mainly depends upon the aperture size of the telescope and divergence.  A smaller beam size leads to greater losses at the receiver due to its higher divergence. On the other hand, while a larger beam size exhibits less divergence, it becomes more susceptible to turbulence. Therefore, achieving an optimal beam size is crucial for improved performance. The gain of the transmitter beam \cite{NEL}, $G_t$ is given in \eqref{eqgt}.
 
 ${\Theta_{B}}$ is related to wavelength $\lambda$ and beam size $W_{0}$ as $\Theta_{B}=\frac{\lambda}{\pi W_{0}}$.
 
 The telescopes used are of Cassegrain-type architecture and the beams used are Gaussian type for both the transmitter and receiver side. Due to the obstruction of the secondary mirror, the overall transmittance of the telescope is given by \,\cite{Gt}

 \begin{equation}
   \eta_{t} = \frac{2}{\alpha^2}\left(e^{-\alpha^2} - e^{-\alpha^2\gamma^2}\right)^2
 \end{equation}
  where $\gamma = \frac{b}{R} $. $b$ is the secondary radius and $R$ is the primary radius of the telescope.

  $\alpha = \frac{R}{\omega}$. This ratio corresponds to the maximum fractional transmission by the telescope aperture. $\omega$ is the beam radius at the transmitter end.\\

\subsection{Path loss:}
Here, the total path loss is categorized into two parts: one due to the satellite's distance from the Earth's surface, known as free space path loss, and the other arising from atmospheric conditions.
The Earth's orbits are categorized into three distinct groups based on their altitudes, each serving specific purposes. The Geostationary Orbit (GEO) resides farthest from the Earth's surface, situated at an altitude of approximately 36,000 kilometers. In contrast, the Low Earth Orbit (LEO) is in the closest proximity to Earth, ranging from 200 to 2,000 kilometers in altitude, and it boasts the highest relative speed among these orbital categories. The Medium Earth Orbit falls between these two extremes \,\cite{progress_2017}.

Given its proximity to Earth, LEO experiences minimal diffraction loss, making it particularly advantageous for quantum key distribution (QKD) experiments. However, orbits with slower speeds, such as the Middle Earth Orbit (MEO) or GEO, offer the benefit of maintaining a continuous link for extended periods, allowing for prolonged QKD operations. Nonetheless, it's worth noting that MEO and GEO orbits come with their own challenges, including higher radiation levels and increased propagation loss.
 
Free space propagation loss \cite{Pathloss} in dB is $\log_{10} L_r$ where $L_r$ is given in \eqref{eqgt}.

\subsubsection{Atmospheric loss:}
 The atmosphere plays a crucial role in the quantum beam. The cause of atmospheric losses on the quantum beam can be divided into two parts. Such as static components and dynamic components.
It can be written as 
\begin{equation}
    \eta_{atm} = \eta_{atte}\eta_{tur}\,,
\end{equation}

$\eta_{atte}$ is due to the static components of the atmosphere and $\eta_{tur}$ \cite{Sharma2019}is due to the turbulence i.e dynamic components.\\
The Static components constitute various types of gases, water vapors, and dust particles that constitute the static atmosphere. These atmospheric elements absorb energy from the photons and this phenomenon is known as absorption. It depends on the wavelength. Photon collisions with air particles cause scattering. The dynamic components are discussed in the next subsection \ref{Turbulence effects}.

Considering a collimated beam of initial beam intensity $I(0)$, after a path of length L, the atmospheric transmittance $\eta_{atte}$ related to atmospheric attenuation may be described by Beer’s law as \,\cite{NEL}

\begin{equation}
    \eta_{atte} = \frac{I(L)}{I(0)} = e^{-\beta_{ext}(h,\lambda)L}\,,
\end{equation}
where $\beta$ is the extinction coefficient and $h$ is the altitude.

\begin{figure}[!h]
    \centering
    \includegraphics[width=1.0\columnwidth]{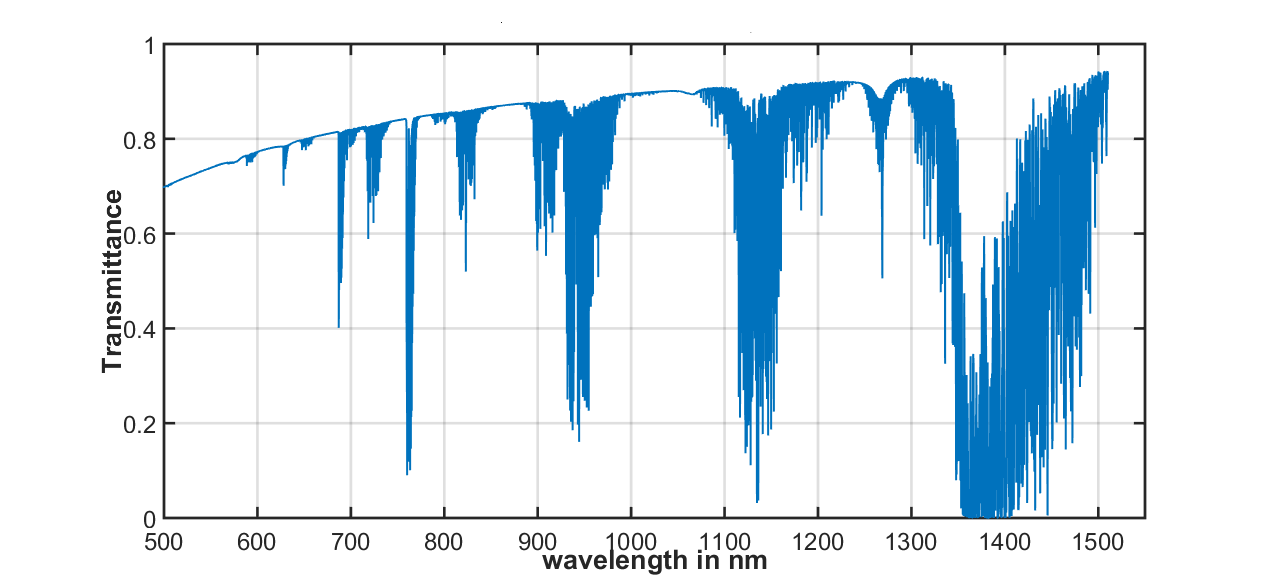}
    \caption{Simulated atmospheric transmittance at the IAO Hanle, a rural location. The X-axis represents the wavelength in nm, ranging from 500 to 1500, while the Y-axis represents the transmittance.}
    \label{fig:IAOHanleTransmttance}
\end{figure}

A number of atmospheric radiative transfer software simulation packages have been developed to describe the effects of atmospheric absorption and scattering under different atmospheric conditions and over a large wavelength range. A popular computer program called MODTRAN models the atmosphere in the range of 100,000 nm (Far-IR) to 200 nm (UV). MODTRAN 6, is being employed in our simulation and offers a spectral resolution of $0.2 cm^{-1}$ \,\cite{MODTRAN}. The transmittance of IAO Hanle is found and given in Fig.\ref{fig:IAOHanleTransmttance} The value of the atmospheric constituents is taken from GIOVANNI open source for the year 2015 from Jan to Dec. Giovanni stands for GES-DISC Interactive Online Visualization and Analysis Infrastructure. It is a web interface provided by NASA that allows users to explore gridded data obtained from various satellite and surface measurements \,\cite{GIOVANNI}. Table \ref{tab:IAOHanleModtran} displays the parameters relevant to the MODTRAN plot.

 \begin{center}
\begin{table}
\centering
\begin{tabular}{|c|c|}
\hline 
Atmospheric parameter & IAO  \\
\hline 
$H_{2}O$ ($g/cm^2$) & 0.0865\\
$CO_{2}$ (ppm) & 390\\
$O_{3}$ ($g/cm^2$) & 265 \\
Aerosol RH & 20 \\ 
Visibility(km) & 23\\
Rain (Mm/hr) & 0.128 \\
Climate & Mid-latitude winter \\
Model & Rural \\
Temperature (kelvin) & 282 \\
Altitude (meter) & 4488 \\
Reflectance & 0.3
\tabularnewline
\hline 
\end{tabular}
\caption{Atmospheric parameters considered for MODTRAN simulations for IAO Hanle.}
\label{tab:IAOHanleModtran}
\end{table}
\par\end{center}

Some commercially available laser systems such as those operating at 532 nm, 780 nm, 810 nm, 1060 nm, and 1550 nm, etc. have distinct transmission windows within this spectrum, where optical transmission encounters minimal loss. In general, higher wavelengths tend to offer improved transmission, but it's essential to consider other factors such as diffraction, as well as the capabilities of sources and detectors when determining the most suitable wavelength choice. We use 810 nm as the main quantum signal beam and 532 nm and 1550 nm as the beacon beam for uplink and downlink respectively. 

\subsubsection{Turbulence}
\label{Turbulence effects}
The structure constant $C_{n}^2$ is a common indicator of wave propagation over random media such as the atmosphere.  It quantifies the variation in refractive index, denoted by $n$, caused by irregularities within the medium. To model the atmosphere effectively, it is essential to comprehend the altitude-dependent characteristics of the structure constant and its sensitivity to weather conditions. Accurately modeling the atmosphere poses a challenge, and different empirical and parametric theories have been developed to describe the altitude-dependent variations in the structure constant. There are various types of atmospheric models used, such as the Hufnagel-Valley (H-V) model, Polynomial Regression (PR) model, and  Submarine Laser Communication (SLC) model etc. Out of these, the  H-V model is widely utilized and is given by \,\cite{hv_2019} 
\begin{equation}
    C^2_{n}(h)=0.00594 \bigg(\frac{v}{27}\bigg)^2 ( 10^{-5}\times h)^{10}\times e^{-10^{-3}h} 
    + \ 2.7 \times 10^{-16} e^{-\frac{2h}{3\times10^3}}
    +\ A_{0}\ e^{-10^{-2}h}\,.
\end{equation}
Here $h$ represents the altitude in meters relative to the specified location and the user is required to set the values of the parameters $v$ and $A_{0}$ as indicated below.

The ground level turbulence strength is represen ted by $A_{0}$ in units of $m^{-2/3}$, while the root mean square (rms) wind speed at high altitude is denoted as $v$ in meters per second. The H-V model provides a convenient way to adjust the profile of $C_{n}^2$ by modifying the values of the parameters $A_{0}$ and $v$. 
Standard value, for $ A_{0} = 1.7 \times 10^{-14} m^{-2/3}$ and $v$ = 21 m/s \,\cite{army}.

An essential parameter for defining the turbulence and the nature of the wavefront propagating in the atmosphere is the atmospheric coherence length.

\begin{equation}
    r_{0} = \left(0.42k^2\sec\theta  \int_{h_{0}}^{Z} C_{n}^2(h) \,dh\right)^{-\frac{3}{5}} \,,
\end{equation}

One can also compute $r_0$ using the `seeing parameter'. We have used this method to calculate $r_0$ for different Indian ground station locations \cite{DevasthalR_0}. When there is no turbulence present, the parameter $r_{0}$ represents the diameter of an equivalent aperture where the resolution of a telescope is approximately diffraction-limited. A larger value of $r_{0}$ implies that turbulence has a lesser impact on the propagation of the beam and vice versa. Since $r_{0}$ varies with the wavelength according to $\lambda^{6/5}$, longer operating wavelengths result in larger values of the $r_{0}$. This suggests that at longer wavelengths, the effect of turbulence on the wavefront is less severe.

One more way to express the strength of turbulence is $\frac{D}{r_{0}}$. This is a ratio of the diameter of the beam from the telescope aperture to the atmosphere's coherence length. The higher the value, the more turbulated the beam. As the value for  $\frac{D}{r_{0}}$ increases higher order wavefront distortion will accumulate \,\cite{roddier1999adaptive}. Hence objective is to have minimum  $\frac{D}{r_{0}}$. But its value shouldn't be $<1$. Though no wavefront distortion will be there but optical beam will go through a beam wander phenomenon which makes it difficult to detect the beam at the receiver. 

\subsubsection{Beam divergence}

Consider symmetric Gaussian beam propagation. In the absence of atmospheric turbulence, the intensity profile of the optical wave at the receiver in free space is given by: \,\cite{Andrews:95}
\begin{equation}
   I_0(r,L)= \frac{W_{0}}{W^{2}} exp(- \frac{2r^{2}}{W^2})\,,
\end{equation}
Here, $W_{0}$ represents the initial beam radius at the starting point where $L=0$, $L$ is the distance between the transmitter and the receiver. The $W$ denotes the diffractive beam radius at the receiver, while $r$ represents a vector perpendicular to the optical beam axis.

The average intensity of the optical wave received at the receiver under the influence of turbulence is given by
\begin{equation}
   \langle I(r,L) \rangle = \frac{W_{0}}{W_{e}^{2}} \exp\left(- \frac{2r^{2}}{W_{e}^2}\right) \,,
\end{equation}
where $W_{e}$ is the effective spot size of the Gaussian beam when atmospheric turbulence is present.

In order to measure the extent of beam spreading, the average effective beam waist is used to describe it as \,\cite{NEL}
\begin{equation}
   W_{e}^2 = W^2 (1+T)\,,
\end{equation}

The parameter $T$ characterizes the additional beam spreading caused by turbulence, which is dependent on the horizontal, uplink, and downlink beam paths as well as the turbulence intensity.

An uplink with a slant route and an angle $\theta$ from the zenith, T can be expressed as follows:
\begin{equation}
   T=4.35 \chi^{5/6} k^{7/6} L^{5/6} \sec^{11/6}\theta \int_{h_{0}}^{h_{0}+L}C_{n}^2(h)(1- \frac{h-h_{0}}{L})^{5/3}dh \,,
\end{equation}
 
where $k$ is wave number, $h_{0}$ is altitude of the observer, $\chi = \frac{2L}{kW^2}$ is amplitude change due to diffraction.

\subsubsection{Beam motion and jitter}
When there's no turbulence, we can express the received signal as $I(r, L)$. $L$  represents the propagation distance.  However, in the presence of turbulence, the average of the received signal becomes $\langle I(r,L) \rangle$. It follows a probability density function denoted as $p(I)$ and its beta distribution  $p(I)$ is given by \,\cite{toyoshima_optimum_2002}

\begin{equation}
    p(I) = \beta I^{(\beta -1)} \,,
\end{equation}
for $0\leq I \leq 1$, 
$\overline I = \frac{\beta}{\beta +1}$
where $I$ is the normalized intensity, $\overline I$ is the average value, and 
\begin{equation}
   \beta = \frac{\Theta_B^2}{4\sigma_j^2}\,,
\end{equation}

The variance $\sigma_j^2$ is made up of the variance of atmospheric-induced and transmitter-induced pointing errors. The angle of arrival perturbation has a Gaussian distribution described by a zero mean and variance along the x and y axes as

\begin{equation}
   \sigma_j^2 =0.182 \left(\frac{D}{r_{0}}\right) ^{\frac{5}{3}} \left(\frac{\lambda}{D}\right)^2\,,
\end{equation}

Though more challenging to measure, the transmitter-induced jitter is often negligible in comparison to atmospheric effects. From Eq 19. it can be concluded that either the beam jitter must be reduced or the angular beamwidth must be raised to attain a high value of $\beta$. In practice, one is frequently constrained to raising the beam divergence in order to get a certain value of $\beta$ because one has little control over the atmospheric turbulence. However, the square of the beam divergence has an inverse relationship with the received power. Therefore, there is an optimum value of $\beta$ that is dependent on the atmospheric conditions.  

\subsubsection{Atmospheric Scintillation}
The log-normal distribution for the Probability Density Function (PDF) of the normalized received intensity, denoted as $S$, is induced by scintillation. It is expressed as follows
\cite{andrews_laser_2001}.
\begin{equation}
    \sigma_{l}^2= 2.24k^{7/6}\sec^{11/6}\theta \int_{h_{0}}^{L} C_{n}^2(h)h^{5/6}dh\,,
\end{equation}
The variance of $S$ itself is given by
$\sigma_{S}^2= \exp(\sigma_{l}^2)-1$

\subsection{Receiver:}
We considered having a 30 cm aperture size at the satellite receiver which is the maximum size allowed for a boarded satellite\,\cite{JGRarity}. The transmittance is given by Eq.2. The gain of the receiver telescope is given by \,\cite{Gr} \\
\begin{equation}
   G_{r} = \frac{4\pi A}{\lambda^2} \frac{2}{\alpha^2} \left[e^{-\alpha^2} - e^{-\alpha^2\gamma^2} \right]^2\,,
\end{equation}

where $A = \pi \frac{D^2}{4}$ is the aperture area and D is the telescope diameter,

The pointing loss ($l_{p}$) is set at $2\mu rad$ at the receiver. The pointing loss is given by the formula \,\cite{lp:87}

\begin{equation}
    l_{p}(\theta) = 4\left[\frac{J_{1}(p)}{p}\right]^2\,,
\end{equation}

Where $p = \pi (\frac{D}{\lambda}) \theta$\\
and  $\lambda$ is the wavelength, $\theta$ is the off-axis pointing angle, and $J_{1}$ is the Bessel function of order one.

Hence the complete equation for the received power in the presence of atmospheric turbulence is 
\begin{equation}
P_{r} = P_{t}\eta^{\sec\theta}_{atm} L_{r}\eta_{t}\eta_{r}G_{t}G_{r}IS\,,
\end{equation}

Where $I$ and $S$ are random variables in the above equation. The PDF of the received signal can be represented by the product of the random variable $I$ following a beta distribution and the random variable $S$ following a lognormal distribution. 

 \section{Link budget calculation for beacon beams}

To establish and sustain a free-space optical link, it is essential for the spacecraft's altitude determination and control system to ensure precise pointing accuracy toward the optical ground station. Conversely, a laser beacon tracking system has the capability to furnish accurate ground-based altitude information, thus facilitating communication from the satellite. A laser beacon is transmitted from the optical ground station with a different wavelength from the quantum signal beam. The satellite will also transmit a beacon beam to get the position of the ground station. For the uplink, we use 532 nm, and for the downlink beacon 1550 nm wavelength. For both uplink and downlink beacons, we fix the beam divergence to be $500\mu rad$. This high divergence leads to footprints of the beacon lasers for the best possible tracking at the receiving ends. Another key advantage of choosing high-divergence beacon beams is the low pointing error loss during transmission. The link budget calculation for the beacon beam is also given in Table \ref{tab:LinkbudgetDownlink} for a ground telescope of 15 cm aperture and a satellite telescope of 30 cm aperture,\cite{Suzuki1997CurrentSO}. The calculation for the optimum size of 15 cm is given in section A.4 of the Appendix.

 \begin{center}
\begin{table}[!ht]
\centering
\begin{tabular}{|c|c|c|c|c|}
\hline 
No & Parameter & unit & Uplink  & Downlink \\
\hline 
1 & Tx power ($P_{t}$) & W &  1.0 & 1.0\\
2 & Wavelength ($\lambda$) & nm & 532  & 1550 \\
3 & Tx Beam diameter & mm &  150.0 & 300.0\\
4 & Tx Beam divergence ($2 \Theta_{B}$) & $\mu rad$ &  500 & 500\\
5  & Tx gain  ($ G_{t}$) & dB & 81.07 & 81.07 \\
6 & Tx optics loss ($ \eta_{t}$) & dB & -2.20 & -2.20 \\

\hline
7 & Propagation distance ($L$) & km & 500  & 500 \\
8 & Path loss ($L_{r}$) & dB & -261.48  & -252.16 \\
9 & Atmospheric transmittance ($\eta_{atm}$)=0.73/0.81 & dB & -1.36 & -0.9  \\
10 & Turbulence ($\eta_{tur}$)=0.64/0.96 & dB & -1.88  & -0.18 \\
\hline 

11 & RX antenna diameter & cm & 30.0 & 15.0\\
12 & RX gain ($G_{r}$) & dB &  124.97 & 109.66 \\
13 &RX optics loss ($ \eta_{r}$) & dB & -2.2 & -2.2 \\

\hline 
14 & Total loss & dB & 63.08 & 66.91 \\
\hline

\end{tabular}
\caption{\label{tab:LinkbudgetDownlink}Link budget for uplink 532 nm and downlink 1550 nm beam. Rows 1 to 6 represent parameter for the transmitter telescope. Rows 7 to 10 represent propagation losses. Rows 11 to 13 represent parameters for the receiver telescope.}
\end{table}
\par\end{center}

To summarize the information contained in the various tables, the first Table \ref{tab:LinkBudget} provides the link budget for the signal beam at the IAO Hanle observatory. The atmospheric parameters for the period from January to December 2015 which are sourced from the GIOVANNI database presented in Table \ref{tab:IAOHanleModtran}. Table \ref{tab:LinkbudgetDownlink} presents the link budget for the uplink and downlink beacon beams at IAO Hanle. Finally, Table \ref{tab:LinkBudgetAllIndianLocation} compares the link budget analysis for the signal and beacon beams between IAO Hanle, Aries Nainital, and Mount Abu. For a comprehensive understanding of the parameters employed to compute the link budget for these two observatories, please refer to Table A4 within the Appendix section. The beacon beam at 532 nm is used for uplink and 1550 nm for downlink, whereas the last column for the signal beam is at 810 nm. For the downlink beacon, due to the smaller size of the ground telescope, the loss is always greater compared to the uplink.

\begin{center}
\begin{table}[!ht]
\centering
\begin{tabular}{|c|c|c|c|}
\hline
Location & Uplink beacon loss in dB & Downlink beacon loss in dB & Signal loss in dB\\
\hline 
IAO Hanle & 63.08  & 66.91  & 35.91 \\

\hline
ARIES Nainital & 67.22  & 67.54  & 37.78 \\

\hline
Mount-Abu & 66.26  & 66.93  & 37.19 \\
\hline
\end{tabular}
\caption{Link budget analysis for 3 different locations in India for uplink beacon, downlink beacon and signal beam of wavelength 532 nm, 1550 nm and 810 nm respectively.}
\label{tab:LinkBudgetAllIndianLocation}
\end{table}
\par\end{center}

A conclusion can be made that IAO Hanle is more suitable for establishing a ground-to-satellite communication channel compared to the other two observatories due to less loss in both uplink and downlink for the signal as well as beacon beams. 

\begin{figure}[!h]
    \centering
    \includegraphics[width=0.7\columnwidth]{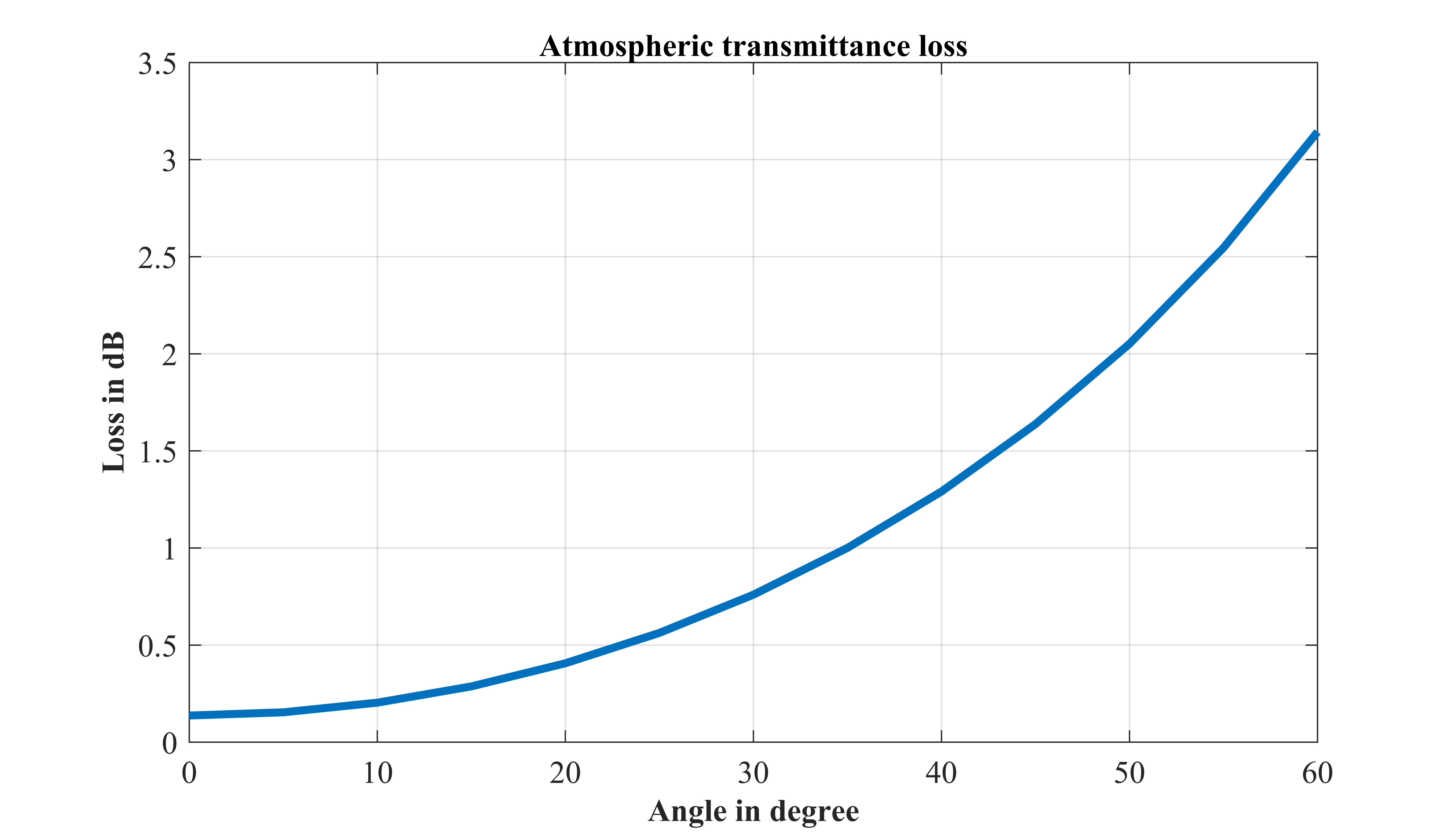}
    \caption{Atmospheric transmission at different zenith angles. X-axis represents the zenith angles in degree. Y-axis represents the loss in dB which will accumulate with increase in angle.}
    \label{fig:TransmittanceWithAngle}
\end{figure}

When the zenith angle is zero degrees, the loss is indicated in Table 4 for the three separate places in India. The air mass will increase when the satellite moves at a greater zenith angle, which is the cause of an increase in loss in dB, as depicted in Fig.\ref{fig:TransmittanceWithAngle}. The plot is given for one of the locations i.e. IAO Hanle. 

The uplink signal loss will rise cumulatively as the satellite passes the vertical direction with increasing zenith angle. The overall cumulative loss is only 5 dB up to $45^{\circ} $. But at $60^{\circ}$, it doubles to 10 dB. Hence, the overall losses at a larger zenith angle are more i.e. the beam passes through the more turbulent region. 

\subsection{Doppler effect assessment}
In the case of a LEO satellite, the apparent velocity relative to the ground exhibits a linear increase as the satellite's elevation angle rises. This scenario involves an analysis of the Doppler effect. As per the Doppler effect, the central frequency of the signal undergoes substantial changes during the data transmission and communication process, primarily due to the rapid motion of LEO satellites. If the satellite receiving end moves away from the transmitting ground end, the signal frequency will decrease, potentially resulting in a higher error rate upon signal reception and recovery.
Normalized Doppler $\frac{\Delta f}{f}$ is given by \,\cite{DoppLEO}

\begin{equation}
    \frac{\Delta f}{f} =\frac{-1}{c} \frac{r_{E}r\sin(\psi(t)- \psi(t_{0}))\cos(\cos^{-1}(\frac{r_{E}}{r}\cos \theta_{max})-\theta_{max}) \omega_{f}(t)}{ \sqrt{r_{E}^2 + r^2 - 2r_{E}r \sin(\psi(t)- \psi(t_{0})) \cos(\cos^{-1}(\frac{r_{E}}{r}\cos \theta_{max})-\theta_{max})}}
\end{equation}
where,
\begin{itemize}
   \item $r_{E}$= radius of earth which is 6400 km
   \item $r$ = position of satellite w.r.t center of earth which is 6900 km
   \item $\psi(t)- \psi(t_{0})$ is satellite visibility window duration which is 8 min. approx.
   \item $\theta_{max}$ is maximum elevation angle which is $90$ degrees.
   \item $\omega_{f}(t)$ angular speed of satellite which is 0.00105 rad/sec
   \item $c$ is the velocity of light. 
   \end{itemize}
For 810 nm signal beam the $\frac{\Delta f}{f}$  comes out to be $1.5\times 10^{-5}$. Correspondingly the frequency shift $\Delta f$  w.r.t the center carrier frequency 380 THz is 0.0057 THz. This shift is considered negligible.
   
\section{Polarization-based quantum communication protocol simulations}
Polarization is a fundamental property of photons given by the plane in which the electric field vector oscillates. Polarization states can be represented as vectors in a two-dimensional Hilbert space\cite{grynberg2010introduction}. Different mutually unbiased bases in this degree of freedom, like \{horizontal (H), vertical (V)\}, \{diagonal (D) and anti-diagonal (A)\} and \{right-circular (R) and left-circular (L)\}, can be used to create bipartite polarization-entangled states. For example, an entangled pair can be represented as

 \begin{equation}
 \psi = \frac{1}{\sqrt{2}}(H_{A}V_{B} + V_{A}H_{B})
 \end{equation}

where subscripts A and B denote two different photons.

Polarization states are used to encode information. To measure the polarization of a photon, polarizing beam splitters and waveplates are used. The measurement outcomes depend on the chosen basis (e.g., H/V basis or D/A basis). In the following section \ref{Atmospheric effects on polarization}, we discuss atmospheric effects on polarization, that can cause error in measurement results. Protocols like BB84 with weak coherent pulse sources use polarization states to generate and share cryptographic keys. Details are given in Sec \ref{WCP QKD}, whereas QKD using an Entangled source is discussed in Sec \ref{EPS QKD}.

\subsection{Effect of atmosphere on polarization degree of freedom}
\label{Atmospheric effects on polarization}
In Free-Space QKD, one of the key challenges is the impact of atmospheric turbulence on the polarization state of photons used for encoding quantum information. As polarized photons propagate through the atmosphere, they are subject to turbulence, which can cause random fluctuations in the refractive index of the air causing an effect of birefringence (See the discussion in \ref{Turbulence effects}). These fluctuations affect the phase of orthogonal polarization components differently, leading to a change in overall polarization state of the photons. This leads to a higher Quantum Bit Error Rate (QBER), as the transmitted polarization state may not match the intended one. These effects of turbulence on polarization state and possible mitigation strategies are as follows:

\textbf{Depolarization:} Turbulence can mix different polarization components, converting pure polarized states into partially or completely depolarized mixed states. Such errors in the real world implementation can be countered using adaptive optics based solution that corrects for any wavefront distortion caused during transmission.

\textbf{Birefringence:} Variations in refractive index along different propagation axes create an effect of birefringence, causing phase shifts between different polarization components. There have been successful demonstration of various active\cite{Xavier_2009ActiveFeedbackwithBeacon,Li18ActiveFeedbackClassicalBeacon} and passive \cite{Chatterjee2023} polarization correction techniques that corrects for any polarization scrambling because of birefringence in free-space or in fiber based systems.

\subsection{Simulated results from QKD utilizing a weak coherent pulse (WCP) source}
\label{WCP QKD}
\begin{figure}[!h]
    \centering
    \includegraphics[trim=50 150 70 200,clip,width=0.8\columnwidth]{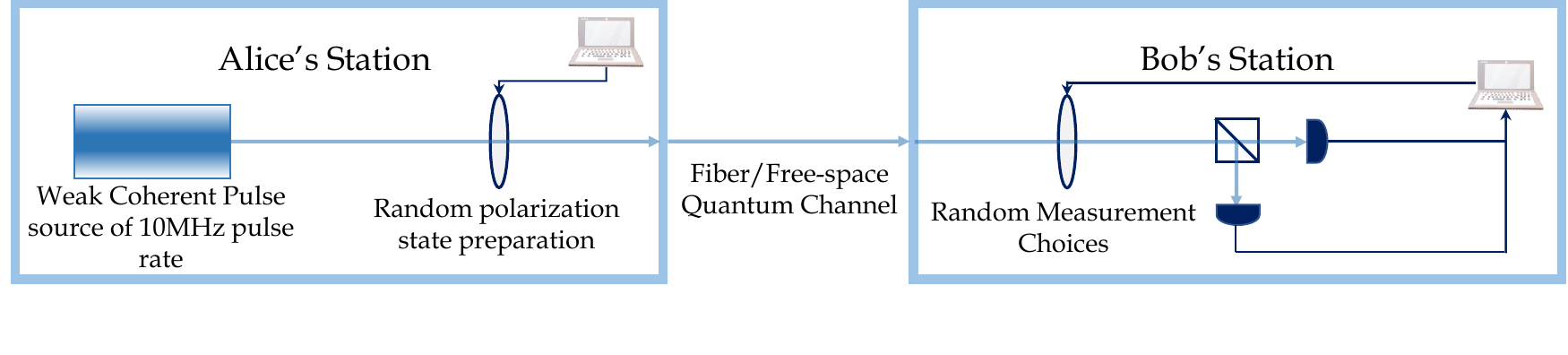}
    \caption{Block diagram of the  BB84 QKD protocol using WCP source with source rate of 10 MHz.}
    \label{fig:Block diagram of BB84 using WCP source}
\end{figure}

QKD can be categorized into two primary types: Discrete Variable QKD (DV-QKD) and Continuous Variable QKD (CV-QKD). DV-QKD systems are further divided into prepare-and-measure protocols and entanglement-based protocols. In a typical prepare-and-measure protocol, Alice encodes each classical bit to a state of a quantum system and sends it to Bob, who performs a specific set of measurements on the incoming signals to retrieve the classical data encoded in their states. In the BB84 protocol, the most widely used prepare-and-measure QKD protocol, each classical bit is encoded into the polarization of a single photon. Alice randomly selects a basis between the horizontal/vertical (Z basis: H, V) and the +45°/-45° (X basis: A and D) and assigns bit values of 0 and 1 to these states. She then sends the chosen photon state to Bob. Bob randomly selects a basis to measure the incoming photon and records the result as a classical bit. After many transmissions, Alice and Bob publicly announce their chosen bases for each photon and discard cases where their bases do not match. They then estimate the error rate from a random subset of their data. If the error rate is low enough, below the security threshold, they apply error correction and privacy amplification to obtain a final shared secret key. A block diagram of BB84 using WCP is given in Fig. \ref{fig:Block diagram of BB84 using WCP source}.

For our simulation, we have assumed a WCP source of 10 MHz pulse repetition rate with a mean photon number($\mu$) of 0.5. Considering this configuration for the source and equivalent sender and receiver optics and detection system, we have applied our link budget analysis to the potential ground station locations under consideration for quantum communication to an LEO satellite.

In the practical implementation of weak coherent pulses in QKD, the quantum link becomes prone towards eavesdropping strategies like photon number splitting attacks. Hence, decoy states are introduced in many experimental demonstrations to prevent potential Eaves from gaining any useful information during the communication \cite{PhysRevLett.94.230504}.

In this work, we have assumed such a decoy state-based system and estimated key rate and QBER considering quantum links from various ground stations in India with fixed source parameters.
The quantum bit error rate for a decoy-state protocol is represented as $E_{\mu}$ and given as \cite{Ma_2005},
\begin{equation}
    E_{\mu} = \frac{ \frac{Y_{0}}{2} + e_{detector}(1-e^{(-\eta \mu)})}{Q_{\mu}}
\end{equation}
where $Y_{0}$ is the dark count of the detector, $Q_{\mu}$ denotes the signal gain, $\eta$ is the detection efficiency and $e_{detector}$ is the probability of erroneous detection.
   
Following the same reference, the asymptotic key rate per pulse was calculated, which has a lower constraint using the decoy state approach. The key rate including the finite key rate analysis for the decoy state-based BB84 QKD protocol is given by
   \begin{equation}
   K \geq q \frac{N_\mu
   }{N_\mu+N_\nu}\Big[ -Q_{\mu}f(E_{\mu})H_{2}(E_{\mu})+Q_{1}\{1-H_{2}(E_{1})-\frac{Q_\mu\Delta}{N_\mu} \}\Big]\,,
   \end{equation}
 
In this context, the parameters are defined as follows: $q = 1/2$ represents the basis reconciliation factor, $f(E_{\mu})=1.22$ stands for the error correction efficiency for practical error correction codes. The binary entropy function is denoted as $H_{2}(x)$. Finally, $Q_{1}$ and $E_{1}$ correspond to the estimated gain and error rate for single-photon pulses. where $N_{\mu}$ $(N_{\nu})$ is
the number of Bob’s received signal (decoy) counts.
The security parameter $\Delta$ is given by
\begin{equation}
    2\log_{2} \frac{1}{[2(\epsilon-\overline{\epsilon}-\epsilon_{EC}) + 7\sqrt{N\log_{2}[\frac{2}{\overline{\epsilon}-\overline{\epsilon}'}]}]}\,,
\end{equation}
where $N$ is the length of the raw key, 

The QBER and key rate calculations for three different observatories are given in Table.\ref{tab: QBER and key rate calculations for 3 different location in India} 

\begin{center}
\begin{table}[!ht]
\centering
\begin{tabular}{|p{2.0cm}|p{2.0cm}|p{2.0cm}|p{2.0cm}|}
\hline
Observatories & Loss (dB) & QBER for WCP in \%  & Keyrates for WCP in bps \\
\hline 
IAO Hanle & 35.91  & 2.29  & 122.48\\

\hline
ARIES Nainital & 37.78  & 2.96  & 73.55\\

\hline
Mount-Abu & 37.19  & 2.72  & 86.63\\
\hline
\end{tabular}
\caption{Estimated QBER and keyrate for 3 different locations in India, considering WCP source-based decoy state QKD protocol. The key rate calculated for the asymptotic limit as given in \cite{Bourgoin_2014}.}
\label{tab: QBER and key rate calculations for 3 different location in India}
\end{table}
\par\end{center}
It can be concluded that all three observatories have QBER less than the information-theoretic security threshold of 11\% meant for BB84 protocol. 

\subsection{Results from QKD simulations using an Entangled photon source}
\label{EPS QKD}
\begin{figure}[!h]
    \centering
    \includegraphics[width=0.8\columnwidth]{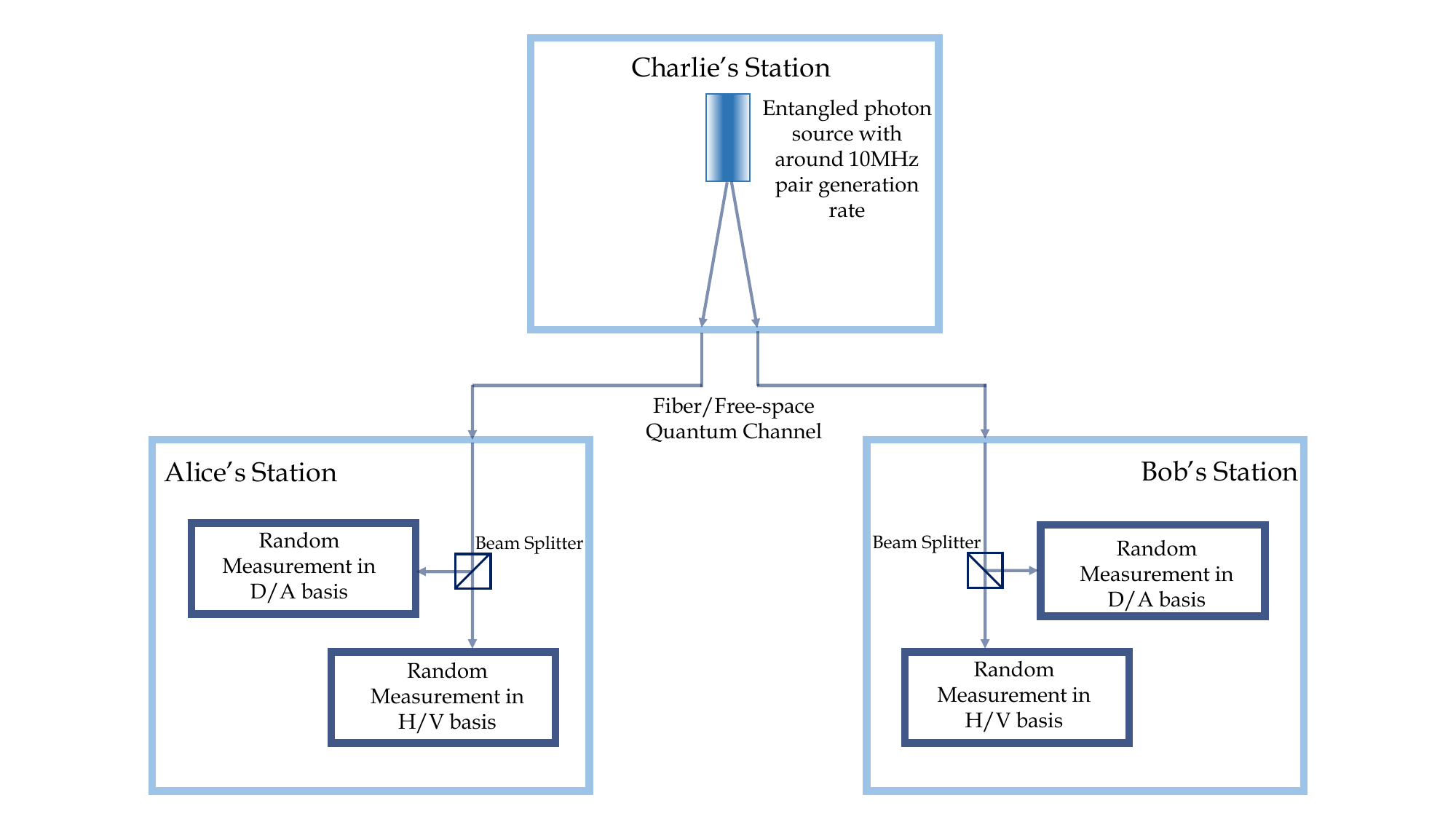}
    \caption{Block diagram of the BBM92 QKD protocol using entangled photon source with 10MHz pair generation rate.}
    \label{fig:QKD using Entangled source}
\end{figure}

Entanglement-based protocols differ from prepare-and-measure systems by eliminating the need for active state encoding. In this work, we consider the BBM92 QKD protocol. Alice and Bob share a source of maximally entangled photon pairs. These photon pairs are entangled in polarization, with one photon belonging to Alice and the other to Bob. Both Alice and Bob independently choose to measure the photons on either a linear or diagonal polarization basis\cite{PhysRevA.61.052304}.
A block diagram of an Entangled source-based QKD protocol is given in Fig. \ref{fig:QKD using Entangled source}.

The transmitter module includes an entangled photon source of 10 MHz pair generation rate. Both Alice and Bob, acting as receivers, have identical polarization-measuring optical setups, each followed by single-photon detectors. A block diagram of BBM92 using entangled photon source is given in Fig.\ref{fig:QKD using Entangled source}

The asymptotic key rate per signal for the BBM92 QKD protocol is given by
 \begin{equation}
  K = \frac{1}{2}Q_\lambda\Big[1-H_{2}(E)-f(E)H_{2}(E) \Big]\,,
 \end{equation}
 
Here, $E$ is the QBER of the signal source. The total number of signals sent by Alice $N$. The term $f(E)H_{2}(E)$ is essentially equal to the measure of the key bits that are transmitted over the public communication channel for error correction purposes. 
$H(x)$ is the binary entropy function. All the parameters used in the equations have been taken from \cite{QKDEntgaled}. The QBER and key rates calculations for Entangled sources are given in Table.\ref{tab:QBER and Keyrate for entangled} for the three observatories.

\begin{center}
\begin{table}[!ht]
\centering
\begin{tabular}{|p{2.0cm}|p{2.0cm}|p{2.0cm}|p{2.0cm}|}
\hline
Observatories & Loss (dB) & QBER for Entangled source in \%  & Keyrates for Entangled source in bps \\
\hline 
IAO Hanle & 35.91  & 5.62  & 14.08\\

\hline
ARIES Nainital & 37.78  & 5.78  & 8.75\\

\hline
Mount-Abu & 37.19  & 5.72  & 10.2\\
\hline
\end{tabular}
\caption{QBER and key rate calculation for 3 different locations in India for Entangled sources.}
\label{tab:QBER and Keyrate for entangled}
\end{table}
\par\end{center}

It can be concluded that all three observatories have QBER less than the information-theoretic security threshold of 11\% meant for the BBM92 protocol.

\begin{figure}[!h]
    \centering
    \includegraphics[width=0.8\columnwidth]{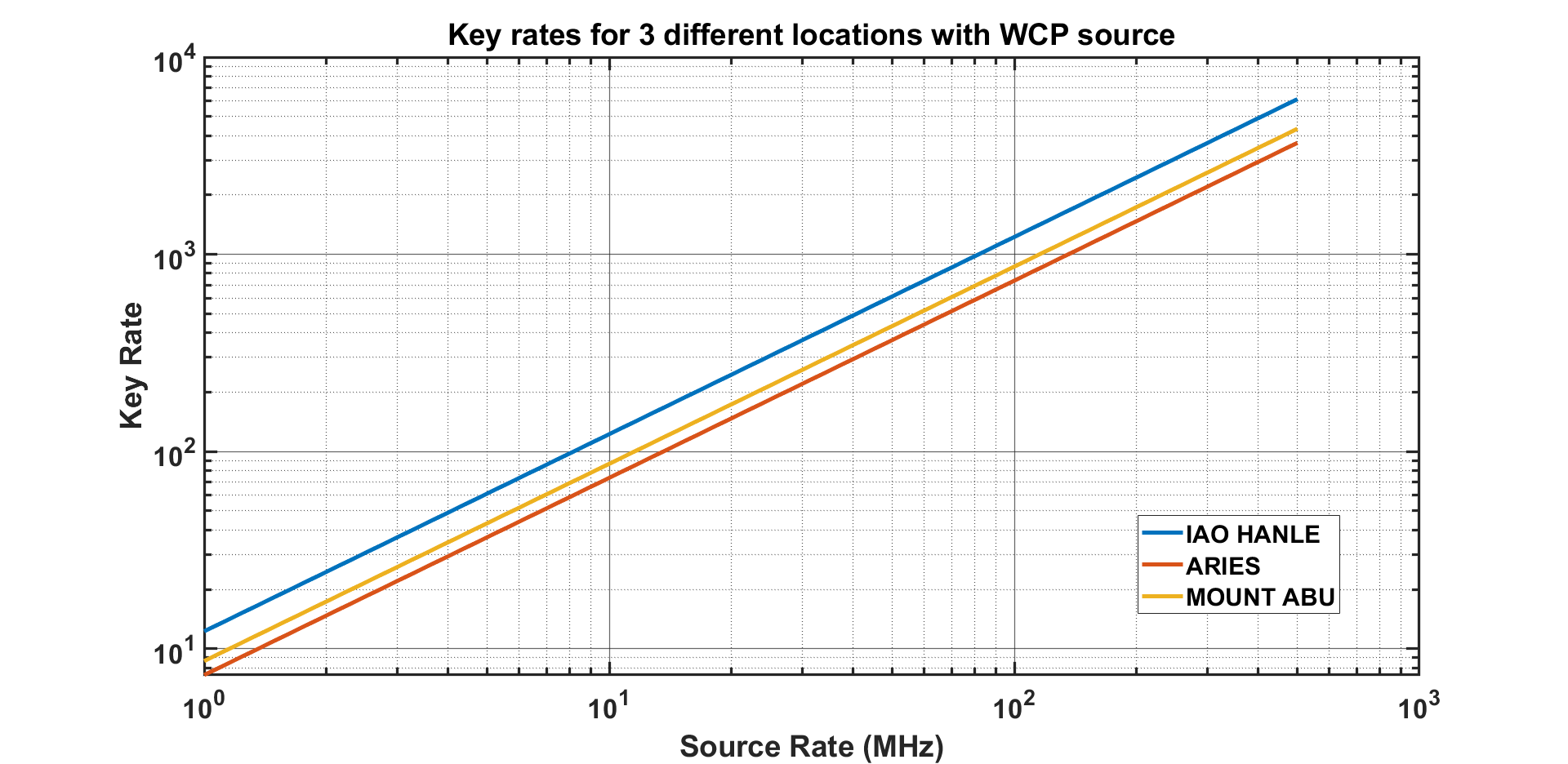}
    \caption{Key rate estimation with BB84 protocol considering WCP source for three different observatories.}
    \label{fig:KeyrateForWCPSource}
\end{figure}

\begin{figure}[!h]
    \centering
    \includegraphics[width=0.8\columnwidth]{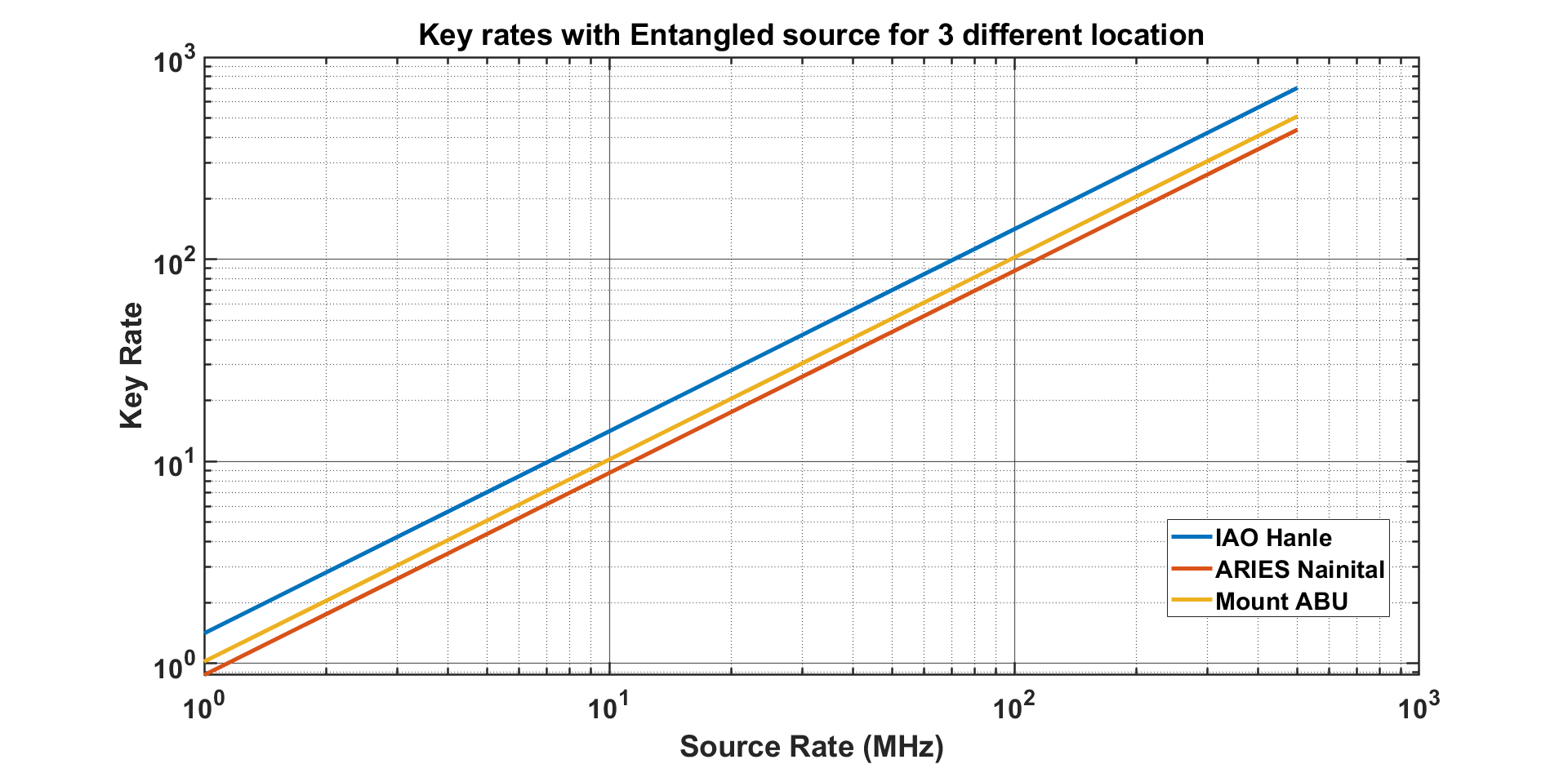}
    \caption{Key rate estimation considering an Entangled source-based QKD for the three potential observatory locations.}
    \label{fig:KeyrateForEntangledSource}
\end{figure}

Fig. \ref{fig:KeyrateForWCPSource} and Fig. \ref{fig:KeyrateForEntangledSource} display the final key rate for both Weak Coherent Pulse (WCP) and Entangled sources for all 3 observatories. The X-axis represents the source rate of the transmitter, while the Y-axis represents the final key rate. The graph includes four detectors, indicating the ratio of the detection window to the repetition period (1.0 ns to 10 ns), and Alice's detection efficiency of 0.25.


By integrating quantum optics simulations for sources and detectors, we ascertain the length of secure keys for Quantum Key Distribution (QKD). This assessment includes considering uplink scenarios for both Weak Coherent Pulse (WCP) and entangled photon sources, each operating at a source rate of 50 MHz. Quantum Bit Error Ratios (QBERs) and sifted key rates are computed and serve as the basis for determining the extent of secure keys. This analysis encompasses all three observatories in the study and is given in Table \ref{tab:LinkBudget3IndianLocations}.

\begin{center}
\begin{table}[!ht]
\centering
\begin{tabular}{|p{2.0cm}|p{2.0cm}|p{2.0cm}|p{2.0cm}|}
\hline
Observatories & Loss (dB) & WCP (bits/sec) & Entangled (bits/sec)\\
\hline 
IAO Hanle & 35.91  & 122.48  & 14.08 \\

\hline
ARIES Nainital & 37.78  & 73.55  & 8.75\\

\hline
Mount-Abu & 37.19  & 86.63  & 10.2\\
\hline
\end{tabular}
\caption{Link budget analysis for 3 different potential locations as ground stations in India.}
\label{tab:LinkBudget3IndianLocations}
\end{table}
\par\end{center}

From Table.\ref{tab:LinkBudget3IndianLocations}, it can be concluded that WCP sources have higher key rates compared to Entangled sources at all the three observatories. QBER for both types of sources for all three observatories is less than 11\%. The schematic provided for the QKD experiments confirms these results. 

\section{Conclusion}
We have presented simulation results analyzing the link performance at three different locations in India. The source being at the ground station, atmospheric effects become more prominent. Consequently, we compared atmospheric absorption using MODTRAN and calculated turbulence-related losses for all observatories. Our findings reveal that the Indian Astronomical Observatory (IAO) in Hanle exhibits the best performance, with total losses of approximately 36 dB. It's important to note that due to turbulence, the beam also experiences increased divergence. Therefore, we have opted for an optimal aperture size of 15 cm, as going smaller results in higher losses, while going larger leads to a more turbulent beam, which in turn reduces the Strehl ratio at the detector. To further enhance performance, additional technologies like Adaptive Optics will be employed.
   
\section{Declarations}
\subsection{Ethical Approval and Consent to participate}
The authors US and SRB  have followed the research ethics guidelines during the entire course of the work. The work has been completed with the full consent to participation from both the contributing authors.
\subsection{Consent for publication}
All the contributing authors (US and SRB) consent to the publication of this work in the journal. Both authors contributed to the writing of the manuscript. No other authors are involved. 
\subsection{Availability of supporting data}
The data generated during the project has been included in the manuscript. The data pertaining to this work is not available online. Further data can be provided by the authors on request.
\subsection{Competing interests/Authors' contributions}
The authors of this work are not aware of any competing interests or conflicts regarding this work. The authors thank Animesh Sinha Roy for his technical assistance with the quantum signal and final key rate analysis. SRB thanks Saumya Ranjan Behera, Melvee George, and Mehak Layal for proofreading the manuscript. The authors thank Kallol Sen for his technical assistance.
\subsection{Funding}
The work has been provided by the Indian Space Research Organization (ISRO) through the QUEST (Quantum Experiments using Satellite Technology) research grant.

\clearpage

\begin{appendices}

\section{Quantum communication in Canary Island:}\label{secA1}

An experimental evaluation was conducted to assess the practicality of a satellite-based global quantum key distribution (QKD) system. The experiment involved a free-space QKD setup over a real distance of 144 kilometers \,\cite{144km}. The Canary Islands of La Palma and Tenerife were selected as the testing locations, with both sites equipped with transmitter and receiver units positioned at an altitude of 2500 meters. The transmission of attenuated laser pulses from the compact and portable transmitter unit to the receiver was facilitated using a 15-cm optical telescope.

 From \cite{144km}, Table \ref{tab:Parameter used to calculate loss between two islands} gives the total specifications in detail for the two islands' communication. 

\begin{center}
\begin{table}[!ht]
\centering
\begin{tabular}{ |c|c| } 
 \hline
 Nordic Optical Telescope (NOT) transmitter & 150 mm  \\ 
 NOT altitude & 2381 m  \\ 
 Optical Ground Station (OGS) receiver & 1 m \\ 
 OGS altitude & 2393 m \\
 Link distance & 143.6 km  \\ 
 Quantum wavelength & 850 nm  \\ 
 Beacon wavelength & 532 nm  \\ 
 $C_{n}^2$ value by H/V model & $1.7\times10^{-14}$  \\ 
 Fried parameter $r_{0}$ & 5cm (at 500 nm wavelength) \\
 \hline
\end{tabular}
\caption{Parameters used to calculate loss between La Palma and Tenerife island.}
\label{tab:Parameter used to calculate loss between two islands}
\end{table}
\end{center}

End-to-end transmission losses $L_{ee}$ were determined by comparing the intensity before the transmitter lens and after passing through the optics of the OGS telescope at the Coude focus. Comparable optical power meters were used for this comparison. These losses encompass four distinct processes, all contributing to $L_{ee}$.

The total transmission losses $L_{ee}$ can be broken down into four components:

$L_{0}$: Beam spreading loss caused by diffraction, which exists even in a vacuum.

$L_{A}$: Atmospheric losses resulting from scattering and absorption by air molecules.

$L_{T}$: Turbulent atmospheric losses caused by beam spreading due to turbulence.

$L_{I}$: Losses due to imperfections in the optical components.

Additionally, the attenuation introduced by the transmitter and receiver telescope optics up to the Coude focus accounted for an approximate attenuation of 4 dB.

The following Table \ref{tab:Different types of losses between two different wavelengths of two islands} presents the measured parameters during a specific nighttime observation. The absorption- and scattering-induced losses, denoted as $L_{A}$, for light at wavelengths 532 nm and 850 nm appear to range between 6 dB and 18 dB and 10 dB and 19 dB, respectively.
\begin{center}
\begin{table}[!ht]
\centering
\begin{tabular}{ |c|c|c|c|c|c| } 
 \hline
 $\lambda$ in nm & $\Theta_{t}$ in $\mu$ radian & $W_{eff}$ in m & $L_{A}$ in dB & $L_{T}$ in dB & $L_{ee}$ in dB \\ 
  \hline
 532 & 23.6 & 3.4 & 13.5 & 14.2 & 31.7 \\ 
 \hline
 850 & 39.6 & 5.7 & 18.8 & 18.7 & 41.5 \\ 
 
 \hline
\end{tabular}
\caption{Different types of losses for different wavelengths between the two considered islands.}
\label{tab:Different types of losses between two different wavelengths of two islands}
\end{table}
\end{center}

\subsection{Validation of losses observed in Canary Island:}

The atmospheric mode chosen is the standard Hufnagel-valley model (H/V) with $A =1.7\times 10^{-14}$  and the Fried parameter is chosen 5cm for 500 nm. The $L_{T}$  matches with the table given for a transmitter beam radius of 15 cm for a distance of 144 km in horizontal propagation.
For horizontal links, one can use a semiempirical method to describe and scale the scattering extinction coefficient. For a wavelength of $\lambda= 550 nm$, the visual range $V$ is defined as the distance in (Km) in a horizontal path with constant scattering. On a normal clear day, the visual range can have a value of $V$ of up to 23 kilometers, but on a cloudy day, it can only be as low as 5 kilometers.

The MODTRAN plots the transmittance with respect to the wavelength. For horizontal transmission $T_{h}$ over a path $(L)$ at any altitude $(h)$ , the attenuation Coefficient $T_{h}$ \,\cite{elterman_uv_1968}.

\begin{equation}
T_{h}=e^{-\beta_{ext}(h,\lambda)L}\,.
\end{equation}

This expression is known as Beer-Lambert law which is the ratio of the collimated beam after propagating distance L, $I(L)$  with initial beam intensity $I(0)$ then $\frac{I(L)}{I(0)}=T_{h}$

\begin{figure}[!h]
    \centering
    \includegraphics[scale=0.45]{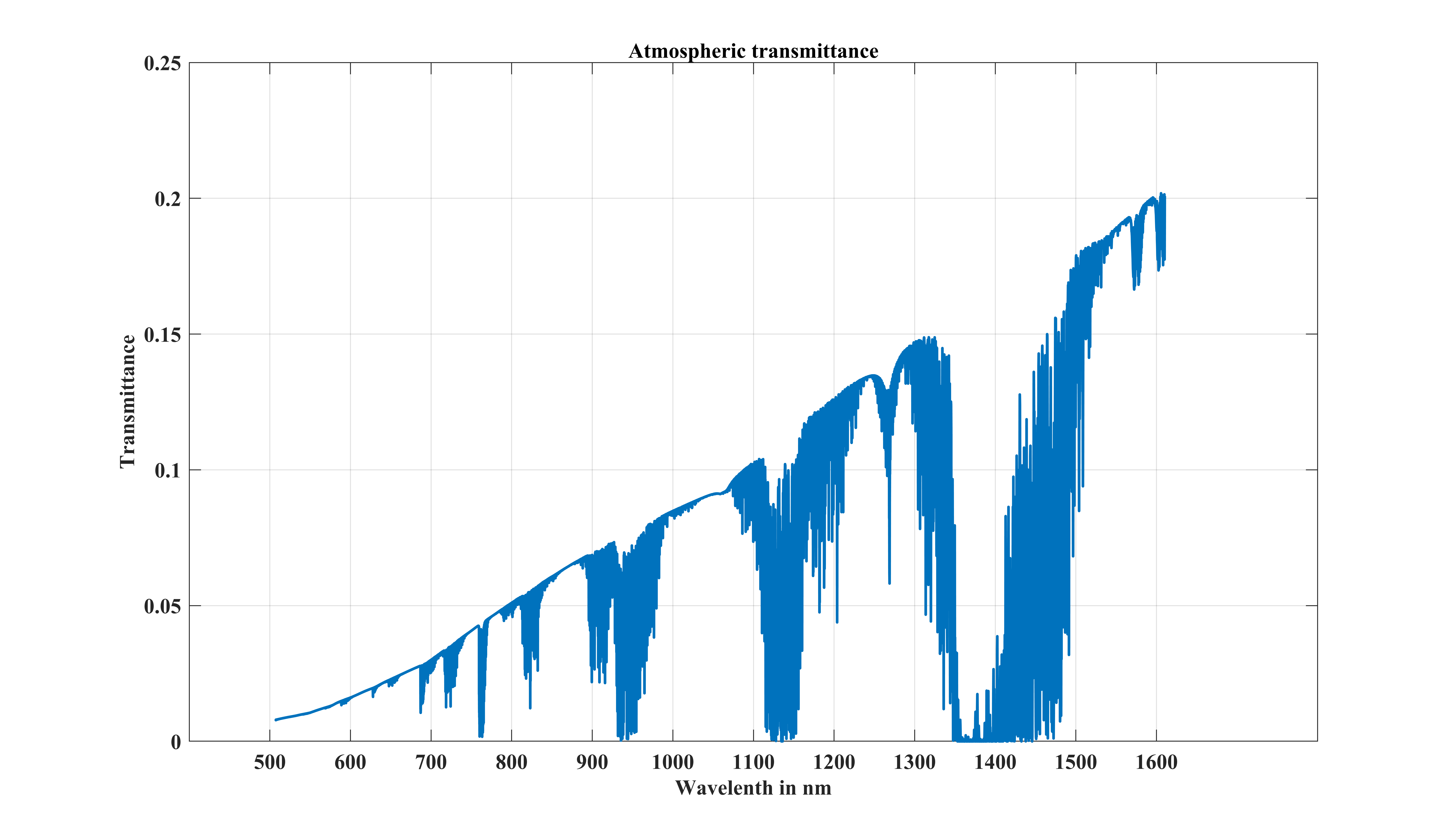}
    \caption{Atmospheric transmittance of Canary islands. X-axis represents wavelength in nm from 500 to 1500. Y-axis represents the transmittance value.}
    \label{fig:TransmittanceCanaryIsland}
\end{figure}

The MODTRAN parameter is given in Table \ref{tab: MODTRAN for Canary}  for the plot of atmospheric transmittance with the wavelength for the Canary Island \,\cite{144km} is given in Fig. \ref{fig:TransmittanceCanaryIsland}. From this plot, it can be concluded that the transmittance at 850 nm matches the loss given in Table \ref{tab:Different types of losses between two different wavelengths of two islands}, since it is within the range given in the ref \cite{144km}.

\begin{center}
\begin{table}[!ht]
\centering
\begin{tabular}{ |c|c| } 
 \hline
 Model & Mid-latitude  \\ 
 Relative Humidity (RH) & 90 \%  \\ 
 Aerosol model & Maritime   \\ 
 Visibility & 5 km  \\ 
 Optical depth & 3.778  \\ 
 Altitudes & 2400 meter  \\ 
 \hline
\end{tabular}
\caption{Parameters used in MODTRAN simulations for Canary Islands.}
\label{tab: MODTRAN for Canary}
\end{table}
\end{center}
\bigskip


\subsection{Verification of Canadian data through simulation:}
\begin{figure}[!h]
    \centering
    \includegraphics[scale=0.5]{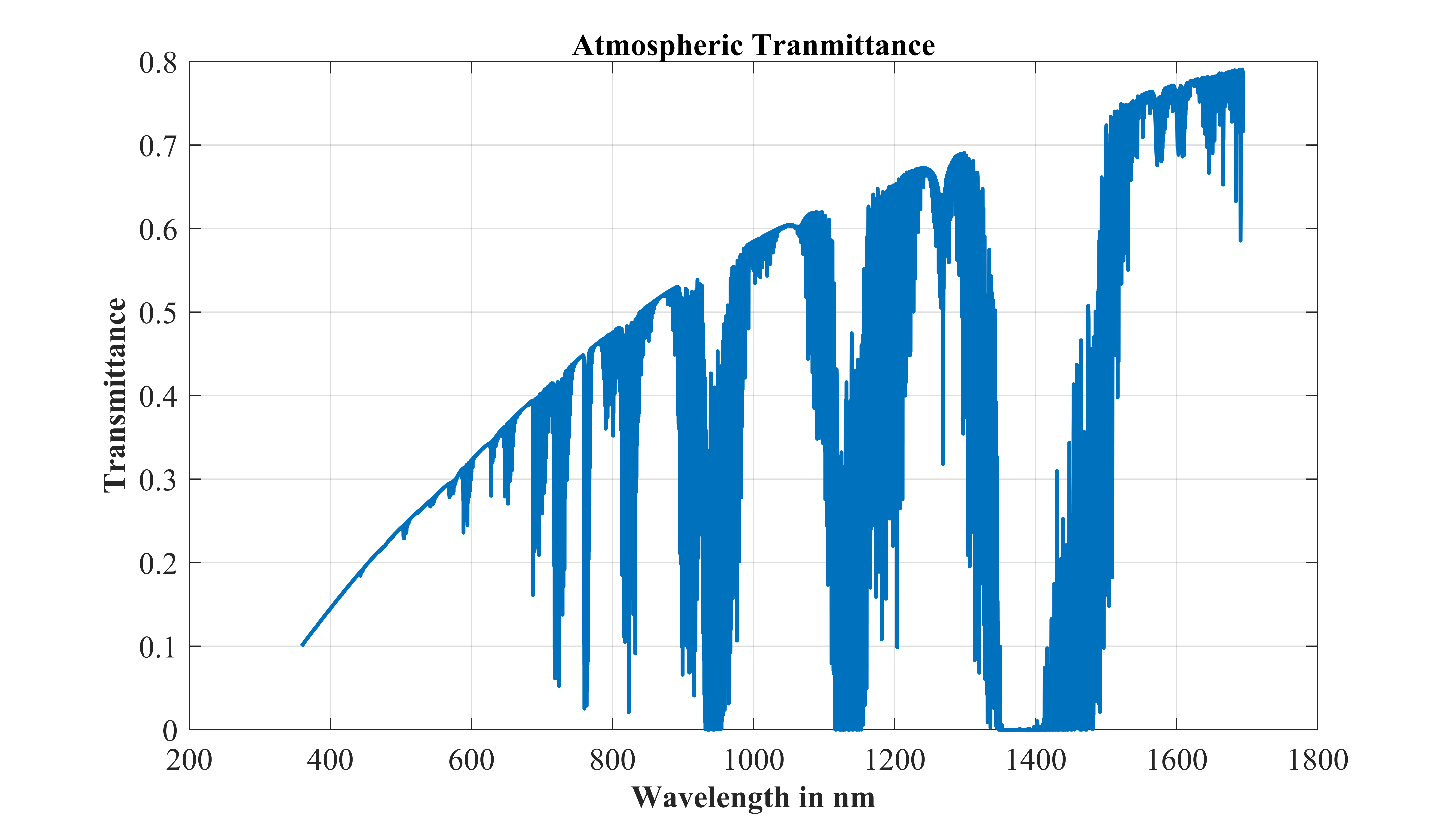}
    \caption{Atmospheric transmittance from the considered Canadian location(Ottawa).}
    \label{fig:TransmittanceCanadianLocations}
\end{figure}

The Canadian Quantum Encryption and Science Satellite mission at the University of Waterloo proposed the uplink as well as downlink QKD experiment for the Ottawa location in Canada for an LEO orbit satellite. This location is situated in the sub-Arctic region with an altitude of 70m, and a sea-level rural atmosphere with a visibility of 5 km \,\cite{PhysRevLett.98.010504}. The plot is given in Fig. \ref{fig:TransmittanceCanadianLocations}.   

\subsection{Modtran parameters for three Indian observatories:}

\begin{figure}[!h]
    \centering
    \includegraphics[width=1.0\columnwidth]{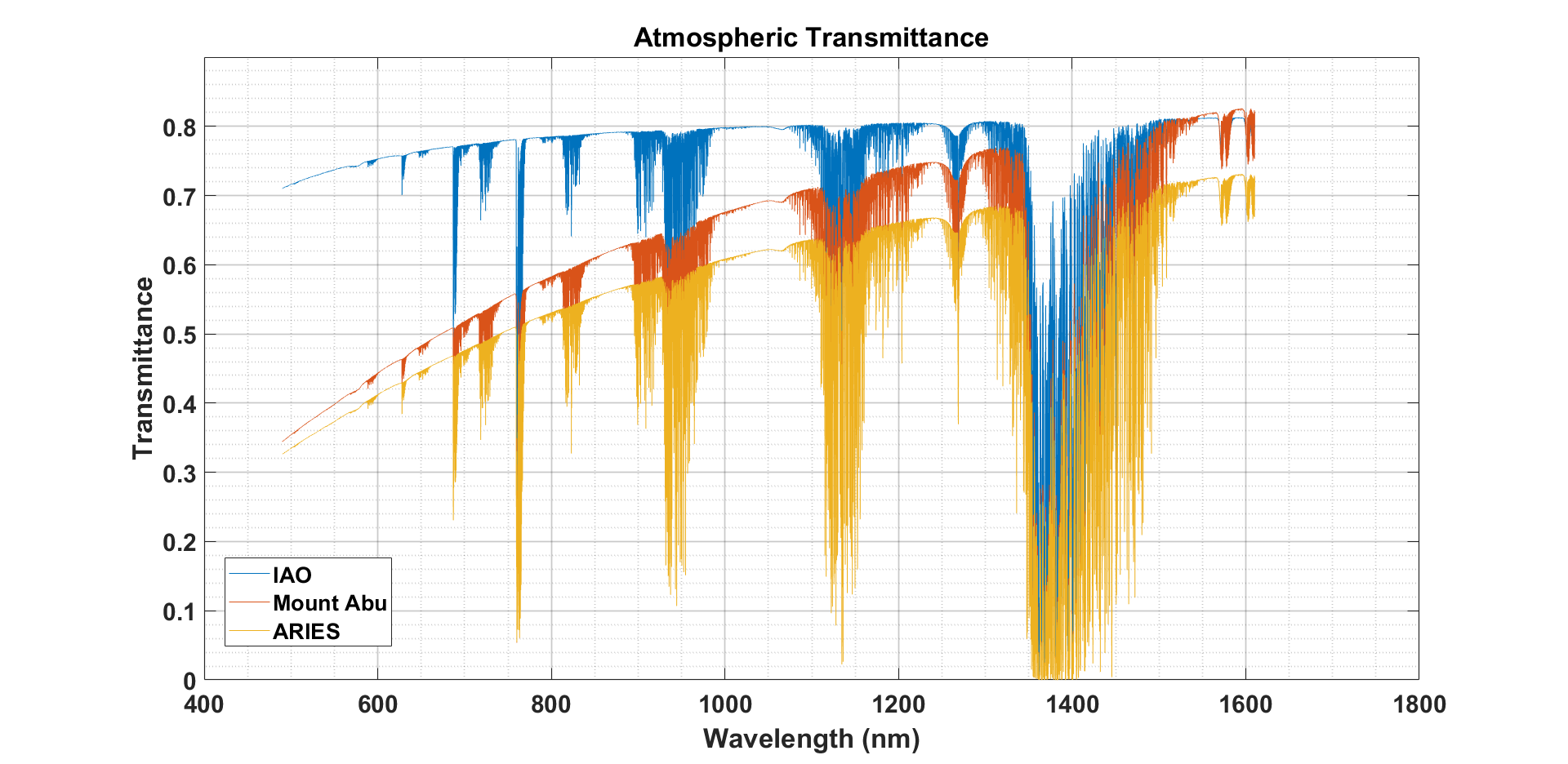}
    \caption{Atmospheric transmittance comparison of three potential observatory locations.}
    \label{fig:TransmittanceComparsion}
\end{figure}

Table \ref{tab:ModtranPlotsIndianLocations} gives the Modtran parameter used for three Indian observatories. Fig.\ref{fig:TransmittanceComparsion} gives a comparison atmospheric transmittance plot for these three observatories.

\begin{center}
\begin{table}
\centering
\begin{tabular}{|c|c|c|c|}
\hline 
Atmospheric parameter & IAO  & Aries & Mount-Abu\\
\hline 
$H_{2}O$ ($g/cm^2$) & 0.0865 & 0.2489 & 0.1732\\
$CO_{2}$ (ppm) & 390 & 322 & 364\\
$O_{3}$ ($g/cm^2$) & 265 & 277 & 270\\
Aerosol RH & 20 & 50 & 65\\ 
Visibility(km) & 23 & 23 & 23\\
Rain (Mm/hr) & 0.128 & 1.29 & 0.82\\
Climate & Mid-latitude winter & Tropical & tropical\\
Model & Rural & Rural dense & Rural\\
Temperature (kelvin) & 282 & 292 & 290\\
Altitude (meter) & 4488 & 1895 & 1220\\
Reflectance & 0.3 & 0.177 & 0.204

\tabularnewline
\hline 
\end{tabular}
\caption{Atmospheric parameters considered in MODTRAN simulations for IAO Hanle, Aries Nainital and Mount-Abu.}
\label{tab:ModtranPlotsIndianLocations}
\end{table}
\par\end{center}

\subsection{Telescope size optimization:}
\begin{figure}[!h]
   \centering
   \includegraphics[width=0.8\columnwidth]{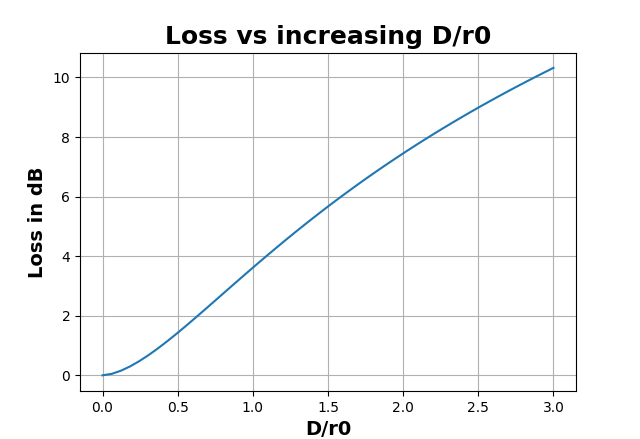}
   \caption{The plot of increasing Strehl ratio loss in dB vs $\frac{D}{r_{0}}$ value.}
   \label{fig:model1}
\end{figure}

The atmosphere plays a major role when a beam in the optical domain propagates through it. Turbulence degrades its quality. The turbulence strength can be measured with Fried parameter $r_{0}$. The main objective is to have a bigger value of the Fried parameter. If D= diameter of the beam from the telescope then a quantity $\frac{D}{r_{0}}$   is a ratio of the diameter of the beam from the telescope to the diameter of the turbulence strength i.e. Fried parameter. The lesser the value is better. Ideally, the value is $\leq 1$. For  $(\frac {D}{r_{0}}) >1 $ indicates that the beam is more turbulent at the focal point, leading to a degradation in beam quality. As this value increases, it results in greater losses at the detector, as illustrated in the Fig.\ref{fig:model1}. However, after $(\frac {D}{r_{0}})>2 $ the degradation of beam starts rapidly \,\cite{NEL}. It can be concluded that $(\frac{D}{r_{0}})$ in between 1 to 2 is safe to operate.

Turbulence and optical aberrations can cause a reduction in the intensity of the beam detected by the satellite. This change is quantified by the Strehl ratio (S), which represents the ratio of the measured on-axis intensity of the detected spot to the intensity of a spot that would be diffraction-limited.

\begin{equation}
    \langle S\rangle = \frac{\langle I(L)\rangle}{\langle I(0)
    \rangle} = \left[1+ \left(\frac{D}{r_{0}}\right)^{5/3}\right]^{-6/5}\,, \ \     
     0<\frac{D}{r_{0}}<\infty
\end{equation}

Which implies that $S$ decreases with increasing $(\frac{D}{r_{0}})$ causing difficulty in detection \,\cite{2023qeyssat}. To calculate attenuation due to turbulence, we use an approximation of the above as given by \cite{Sharma2019}.

 When the ratio $(\frac{D}{r_{0}})$ is less than 1, it leads to the phenomenon of beam wander, which results in instability in the beam. This instability makes it challenging to precisely align the beam with the detector, leading to an increase in pointing errors. Beam wander is given by
 \,\cite{mohammadi2021design}

 \begin{equation}
     \sqrt{\langle r_{c}^2 \rangle} = 0.73Z\sec\zeta\left(\frac{\lambda}{2W_{0}}\right) \left(\frac{2W_{0}}{r_{0}} \right)^{5/6}
 \end{equation}

Where $\langle r_{c}^2 \rangle$ is variance of beam shifting and 
$sec(\zeta)$ is the elevation angle of the beam from the ground station. Further, the angular beam wander can be defined as $\theta_{BW}=\frac{\sqrt{\langle r_{c}^2 \rangle}}{L}$, using which we calculate the loss due to beam wander in our considered uplink scenario \cite{BeamWander1989}.

\subsubsection{Beam divergence under turbulence:}
Optical beams through the atmosphere can be studied by Andrews and Philip’s proposed PDF (probability density function) known as I-K distribution. This PDF is expressed in terms of Bessel functions and can be used to characterize turbulent optical channels over a wide range of operating conditions. Optical field intensity at the receiver is given by 

\begin{equation}
\langle I(x,y,t)\rangle= \frac{P_{t}G_{a}}{4\pi Z^2}\eta_{t}\eta_{r}\eta_{atm}I_{u}\exp \left(- \frac{4(x^2 +y^2)}{\Theta_{B}^2 Z^2}\right)
\end{equation}
\begin{itemize}
\item $\eta _{t,r, atm}$ represents the transmittance of the transmitter, receiver telescopes, and atmosphere, respectively.
\item $P_{t}$ is the transmitted power.
\item $Z$ is the propagation distance.
\item $\Theta_{B}$ is the beam divergence angle.
\item $I_{u}$ is the random variable due to atmosphere scattering.
\end{itemize}
The PDF of $I_{u}$  follows $\beta$  distribution. Where  $\beta= \frac{\Theta_{B}^2}{4\sigma_{jit}^2}$

$\sigma_{jit}^2$ 1-axis 1- jitter variance, which can be represented by
\begin{equation}
\sigma_{jit}^2 = 0.36 \left(\frac{D}{r_{0}} \right)^{5/3} \left(\frac{\lambda}{D} \right)^{2}
\end{equation}

Either the angular beam width or the beam jitter must be decreased in order to attain a high value of $\beta$. In practical scenarios, it is often necessary to increase the beam divergence to achieve a desired value of $\beta$ due to the limited control over atmospheric turbulence-induced beam wander. However, the square of the beam divergence has an inverse relationship with the received power. Therefore, there is an ideal value of $\beta$ that is dependent on the atmosphere. 

A smaller beam diameter will primarily experience beam wander losses, while a larger beam diameter will primarily incur Strehl ratio losses. These losses contribute to the overall total loss. Therefore, the optimal beam size is one that minimizes the combined impact of these two losses, resulting in the lowest total losses.

\begin{figure}[!h]
    \centering
    \includegraphics[width=0.8\columnwidth]{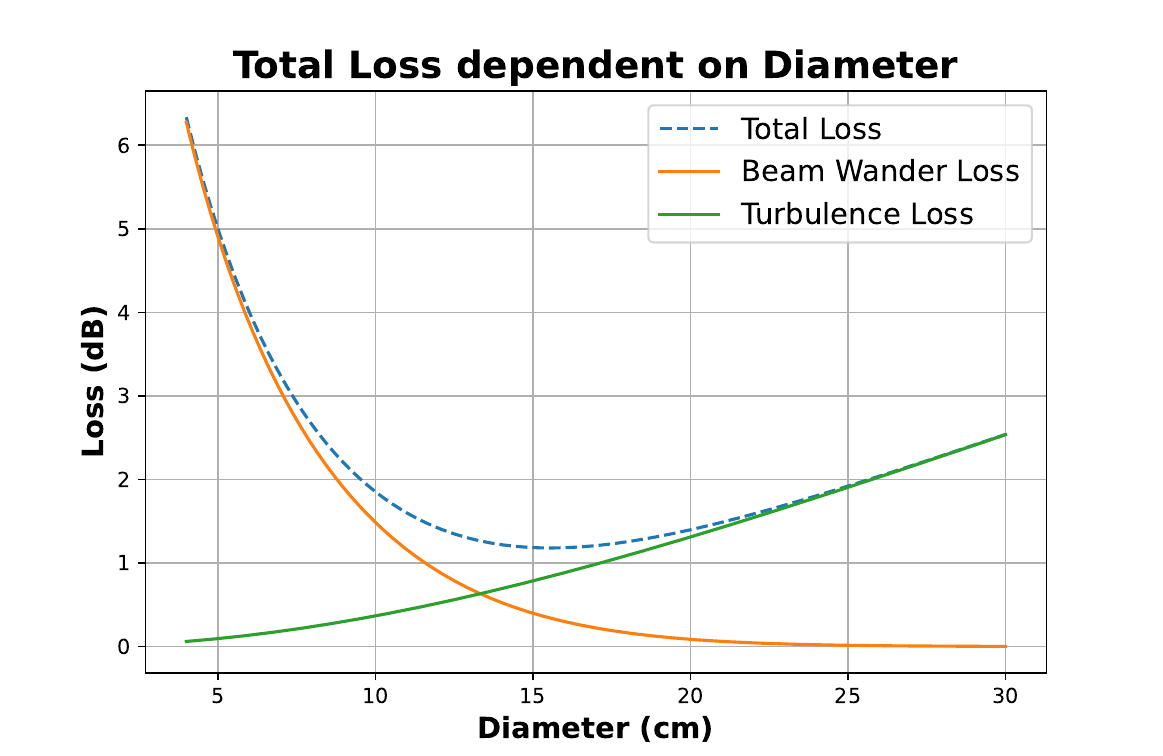}
    \caption{Beam wander and Turbulence losses for different initial beam sizes.}
    \label{fig:LossesWithBeamSize}
\end{figure}

From Fig. \ref{fig:LossesWithBeamSize}. it can be concluded that the beam size 14-15 cm has the minimum losses. Table A5 gives the final loss calculated for IAO Hanle for different beam sizes.

\begin{table}[htbp]
\centering
\resizebox{\textwidth}{!}{
\begin{tabular}{ |c|c|c|c|c|c|c|c| } 
 \hline
 D in cm & 8 cm & 12 cm & 14 cm & 15 cm & 16 cm & 20 cm & 24 cm \\ 
  \hline
 Beam wander loss & 2.38 & 0.87 & 0.52 & 0.40 & 0.30 & 0.08 & 0.02  \\ 
  \hline
 Turbulence loss  & 0.24 & 0.52 & 0.69 & 0.78 & 0.89 & 1.32 & 1.78 \\ 
  \hline
 Total loss dependent on D & 2.62 & 1.39 & 1.22 & 1.18 & 1.19 & 1.4 & 1.80 \\ 
  \hline  

\end{tabular}
}

\caption{Total loss due to turbulence added for different initial beam sizes.}
\end{table}


\end{appendices}

\newpage

\bibliography{Reference}

\end{document}